\documentclass[12pt]{article}

\usepackage{a4}

\begin{document}
\title{Particle-like solutions to higher order curvature
Einstein--Yang-Mills systems in $d$ dimensions}
\author{{\large Y. Brihaye}$^{\ddagger}$,
{\large A. Chakrabarti}$^{\diamond}$
and {\large D. H. Tchrakian}$^{\dagger \star}$ \\ \\
$^{\ddagger}${\small Physique-Math\'ematique, Universite de 
Mons-Hainaut, Mons, Belgium}\\ \\
$^{\diamond}${\small Laboratoire de Physique
Theoriqued, Ecole Polytechnique, Palaiseau, France
}\\ \\
$^{\dagger}${\small Department of
Mathematical Physics, National University of Ireland Maynooth,} \\
{\small Maynooth, Ireland} \\
$^{\star}${\small School of Theoretical Physics -- DIAS, 10 Burlington
Road, Dublin 4, Ireland }}

\date{}
\newcommand{\dd}{\mbox{d}}
\newcommand{\tr}{\mbox{tr}}
\newcommand{\la}{\lambda}
\newcommand{\ka}{\kappa}
\newcommand{\al}{\alpha}
\newcommand{\ga}{\gamma}
\newcommand{\de}{\delta}
\newcommand{\si}{\sigma}
\newcommand{\bomega}{\mbox{\boldmath $\omega$}}
\newcommand{\bsi}{\mbox{\boldmath $\sigma$}}
\newcommand{\bchi}{\mbox{\boldmath $\chi$}}
\newcommand{\bal}{\mbox{\boldmath $\alpha$}}
\newcommand{\bpsi}{\mbox{\boldmath $\psi$}}
\newcommand{\brho}{\mbox{\boldmath $\varrho$}}
\newcommand{\beps}{\mbox{\boldmath $\varepsilon$}}
\newcommand{\bxi}{\mbox{\boldmath $\xi$}}
\newcommand{\bbeta}{\mbox{\boldmath $\beta$}}
\newcommand{\ee}{\end{equation}}
\newcommand{\eea}{\end{eqnarray}}
\newcommand{\be}{\begin{equation}}
\newcommand{\bea}{\begin{eqnarray}}
\newcommand{\ii}{\mbox{i}}
\newcommand{\e}{\mbox{e}}
\newcommand{\pa}{\partial}
\newcommand{\Om}{\Omega}
\newcommand{\vep}{\varepsilon}
\newcommand{\bfph}{{\bf \phi}}
\newcommand{\lm}{\lambda}
\def\theequation{\arabic{equation}}
\renewcommand{\thefootnote}{\fnsymbol{footnote}}
\newcommand{\re}[1]{(\ref{#1})}
\newcommand{\R}{{\rm I \hspace{-0.52ex} R}}
\newcommand{\N}{{\sf N\hspace*{-1.0ex}\rule{0.15ex}%
{1.3ex}\hspace*{1.0ex}}}
\newcommand{\Q}{{\sf Q\hspace*{-1.1ex}\rule{0.15ex}%
{1.5ex}\hspace*{1.1ex}}}
\newcommand{\C}{{\sf C\hspace*{-0.9ex}\rule{0.15ex}%
{1.3ex}\hspace*{0.9ex}}}
\newcommand{\eins}{1\hspace{-0.56ex}{\rm I}}
\renewcommand{\thefootnote}{\arabic{footnote}}

\maketitle
\begin{abstract}
We consider the superposition of the first two members of the
gravitational hierarchy (Einstein plus first Gauss-Bonnet(GB)) interacting
with the superposition of the first two members of the $SO_{(\pm)}(d)$
Yang--Mills hierarchy, in $d$ dimensions. Such systems can occur in the
low energy effective action of string theory. Particle-like solutions
in dimensions $d=6,8$ are constructed respectively. Our results reveal
qualitatively new properties featuring double-valued solutions with
critical behaviour. In this preliminary study, we have restricted
ourselves to one-node solutions.
\end{abstract}
\medskip
\medskip
\newpage

\section{Introduction}
\label{introduction}
Gravitational theories in higher dimensions are of current interest in
the contexts of the AdS/CFT correspondence and of theories with large
and infinite extra dimensions with non-factorisable metrics. Since
non-Abelian gauge fields feature in the low energy effective action of
string theory, it is interesting to study the properties of the
corresponding Einstein--Yang-Mills (EYM) systems. It is our purpose in
the present work, to study higher dimensional EYM systems extended by
higher order terms in both gravitational and YM curvatures.

In particular, we construct static particle-like solutions in the
higher curvature version of Einstein--Yang-Mills systems in
$d$-dimensional spacetimes
in which the $SO(d)$ gauge field takes its values in the chiral
representation $SO(d)_{\pm}$. This is analogous to the $4$ dimensional
model with $SO_{\pm}(4)=SU_{\pm}(2)$ Yang-Mills (YM) field interacting
with Einstein gravity, in the work of Bartnik and McKinnon~\cite{BK}.

In the original system studied in \cite{BK}, the gravitational action
provides the required scaling behaviour needed to balance the scaling of
the static Yang-Mills (YM) system in 3 Euclidean dimensions. This
scaling argument is not rigorous because the gravitational action is not
positive definite by construction, but nonetheless it enables the
construction of particle like solutions. As a result these solutions are
unstable and can be viewed as sphalerons\cite{VG} rather than solitons.
The $(d-1)$ static solutions we construct here are expected to have such
an instability, though in this preliminary work we will not carry out
the corresponding stability analysis. We will restrict to the simplest
model incorporating only up to the second order curvature terms in both
the YM and gravitational sectors.

As it happens, to go to $8\ge d>4$ it is necessary to include the
{\it second} member of the YM hierarchy\cite{T} to provide the requisite
scaling, even in the absence of the gravitational GB term. (Similarly for
$12\ge d>8$ it is necessary to include the {\it thrid} member of the YM
hierarchy.) We will restrict our considerations to the first {\it two}
members of the YM hierarchy, and hence to $d\le 8$. Thus we will consider
systems (a) featuring only Einstein gravity, (b) only GB gravity, and
(c) mixed Einstein and GB gravity, but always with the first {\it two} 
members of the YM hierarchy.

In section \ref{models}, we introduce the models to be studied and in
section \ref{classeqns} we derive the classical equations subject to
our spherically symmetric Ansatz. This consists of the $d$-dimensional
analogue of the $4$-dimensional metric Ansatz employed in \cite{BK,VG},
and the spherically symmetric Ansatz for the YM field in the $SO_{\pm}(d)$
(chiral) representation in $d$-dimensions, analogous with the Ansatz
for the $SU(2)$ YM field in the $SO_{\pm}(4)$ (chiral) representation
in $4$-dimensions used in \cite{BK,VG}. Section \ref{numerics} and its
subsections are devoted to our numerical study of these systems and we
give a summary and discuss our results in section \ref{conclusions}.

\section{The model(s)}
\label{models}
The generic system of interest in this work is the superposed Yang-Mills
hierarchy~\cite{T} interacting with the superposed gravitational
hierarchy consisting of the Einstein and all possible Gauss-Bonnet (GB)
terms in a given spacetime dimension $d$.
\be
\label{einstYM}
{\cal L}\ =\ {\cal L}_{grav}\ +\ {\cal L}_{YM}\ .
\ee
In this preliminary work, we will carry out this construction for 
dimensions $d=6,8$ only.

The definition we use for the superposed gravitational hierarchy is
\be
\label{gravhier}
{\cal L}_{grav}=\sum_{p=1}^{P_1}\ \frac{1}{2p}\ \ka_p\ e\ R_{(p)}\ ,
\ee
in which the $p=1$ term describes the usual Einstein gravity, and the
higher $p$ terms the successive GB contributions. In \re{gravhier}
$e={\mbox det}\ e_{\mu}^a=\sqrt{{\mbox det}\ g_{\mu\nu}}$,
$e_{\mu}^a$ are the vielbeins, and $R_{(p)}$ is the $p$-Ricci scalar. In
terms of the totally antisymmetrised $p$-fold product of the Riemann
tensor $R_{\mu\nu}^{ab}$ yielding the $2p$-form $p$-Riemann tensor
(with the notation $[abc...]$ implying total antisymmetrisation of the
indices $a,b,c,...$)
\be
\label{priemann}
R_{\mu_1\mu_2...\mu_{2p}}^{a_1a_2...a_{2p}}=R_{[\mu_1\mu_2}^{[a_1a_2}
R_{\mu_3\mu_4}^{a_3a_4}...R_{\mu_{2p-1}\mu_{2p}]}^{a_{2p-1}a_{2p}]}\ ,
\ee
we define the $p$-Ricci scalar and the $p$-Ricci tensor as
\bea
R_{(p)}&=&R_{\mu_1\mu_2...\mu_{2p}}^{a_1a_2...a_{2p}}\ e_{a_1}^{\mu_1}
e_{a_2}^{\mu_2}...e_{a_{2p}}^{\mu_{2p}} \label{pricciscalar}\\
R_{(p)}{}_{\mu}^a&=&R_{\mu\mu_1\mu_2...\mu_{2p-1}}^{aa_1a_2...a_{2p-1}}\
e_{a_1}^{\mu_1}e_{a_2}^{\mu_2}...e_{a_{2p-1}}^{\mu_{2p-1}}\ ,
\label{priccitensor}
\eea
Leading to the definitions of the $p$-Einstein tensor
\be
\label{peinstein}
G_{(p)}{}_{\mu}^a=R_{(p)}{}_{\mu}^a\
-\ \frac{1}{2p}\ e_{\mu}^a\ R_{(p)}\ .
\ee
Note that the $p$-Ricci scalar \re{pricciscalar} coincides with the
Euler-Hirzbruch density\footnote{This follows immediately from the
alternative definition for the $p$-Ricci scalar in $d$ dimensions:
\[
\hat R_{(p)}=\vep^{\nu_1\nu_2...\nu_{2p}\mu_1\mu_2...\mu_{d-2p}}
e_{\mu_1}^{a_1}e_{\mu2}^{a_2}...e_{\mu_{d-2p}}^{a_{d-2p}}
\vep^{b_1b_2...b_{2p}a_1a_2...a_{d-2p}}\
R_{\nu_1\nu_2}^{b_1b_2}R_{\nu_3\nu_4}^{b_3b_4}...
R_{\nu_{2p}\nu_{2p}}^{b_{2p}b_{2p}}\ .
\]}
in $d=2p$ (even) dimensions, and hence is trivial. In dimensions
$d\le 2p$ on the other hand,
it vanishes by (anti)symmetry irrespective if $d$ is even or odd. Thus
the upper limit in the summation in \re{gravhier} is $P_1=\frac{d-2}{2}$
in even, and $P_1=\frac{d-1}{2}$ in odd dimensions $d$.

The definition we use for superposed YM hierarchy is
\be
\label{YMhier}
{\cal L}_{YM}=\sum_{p=1}^{P_2}\ \frac{1}{2(2p)!}\ \tau_p\ e\
{\mbox Tr\ }F(2p)^2\ ,
\ee
where $F(2p)$ is the $2p$-form $p$-fold totally antisymmetrised product
of the $SO(d)$ YM curvature $2$-form $F(2)$
\be
\label{2pformYM}
F(2p)\equiv F_{\mu_1\mu_2...\mu_{2p}}=F_{[\mu_1\mu_2}F_{\mu_3\mu_4}...
F_{\mu_{2p-1}\mu_{2p}]}\ .
\ee
Even though the $2p$-form \re{2pformYM} is dual to a total divergence,
namely the divergence of the corresponding Chern-Simons form, the density
\re{YMhier} is never a total divergence since it is the square of one.
But the $2p$-form \re{2pformYM} vanishes by (anti)symmetry for $d<2p$ so
that the upper limit in the summation in \re{YMhier} is $P_2=\frac{d}{2}$
for even $d$ and $P_2=\frac{d-1}{2}$ for odd $d$.

We define the $p$-stress tensor pertaining to each term in \re{YMhier} as
\be
T_{\mu\nu}^{(p)}=
\mbox{Tr}\ F(2p)_{\mu\tau_1\tau_2...\tau_{2p-1}}
F(2p)_{\nu}{}^{\tau_1\tau_2...\tau_{2p-1}}
-\frac{1}{4p}g_{\mu\nu}\ \mbox{Tr}\ F(2p)_{\tau_1\tau_2...\tau_{2p}}
F(2p)^{\tau_1\tau_2...\tau_{2p}}\ .
\label{pstress}
\ee

In the present work, we restrict our attention to $d=6$ and $8$, and in
both cases restrict to only the two terms $p=1$ and $2$, both in the
graviational and the YM hierarchies, even though in these dimensions
higher values of $P$ both in \re{gravhier} and \re{YMhier} can be
accommodated. This restriction is motivated by our desire to study the
simplest nontrivial cases. As for our choice of $d=6$ and $8$, these are
the only even dimensions in which the system consisting of \re{gravhier}
and \re{YMhier} is nontrivial, and our restriction in this preliminary
study is motivated by our desire to maximise the analogy with the study
of Bartnik and McKinnon~\cite{BK} which is in (even) $d=4$. Like in the
model of ~\cite{BK}, we will choose the YM gauge group to be $SO(d)$, with
the YM fields taking their values in the chiral representation
$SO_{\pm}(d)$ (which for $d=4$ reduces to $SU_{\pm}(2)$).

\section{The classical equations}
\label{classeqns}
In $d$ dimensional spacetime, we restrict to static fields that
are spherically symmetric in the $d-1$ spacelike dimensions with the
metric Ansatz
\be
\label{metric}
ds^2=-A(r)dt^2+B(r)^{-1}dr^2+r^2d\Omega_{d-2}^2
\ee
where $r$ is the $(d-2)$ dimensional spacelike radial coordinate and
$d\Omega_{d-2}$ is the $d-2$ dimensional angular volume element.

We take the static spherically symmetric $SO(d)$ YM field in even $d$
to be in the $SO_{(\pm)}(d)$ (chiral) representation
\be
\label{YMsph}
A_0=0\ ,\quad
A_i=\left(\frac{w-1}{r}\right)\Sigma_{ij}^{(\pm)}\hat x_j\ , \quad
\Sigma_{ij}^{(\pm)}=-\frac{1}{4}\left(\frac{1\pm\Gamma_{d+1}}{2}\right)
[\Gamma_i ,\Gamma_j]\ .
\ee
The index $i=1,2,...,d-1$ labels the space dimensions of the static
fields, $\Gamma_{\mu}$ ($\mu=1,2,...,d$) being the gamma matrices in $d$
dimensions. $\Sigma_{ij}^{(\pm)}$ are the generators of the $SO(d-1)$
subgroup of $SO(d)$, the latter being in the chiral representation. We
note that \re{YMsph} is not the most general spherically symmetric Ansatz 
for a $SO(d)$ YM connection and is effectively that of a $SO(d-1)$ field. 
In this sense we could regard our gauge fields to be $SO(d-1)$ rather than
$SO(d)$ fields. In the low dimensional case $d=4$, the generators
$\Sigma_{\mu\nu}^{(\pm)}$ reduce to the those of $SU_+(2)$ and $SU_-(2)$,
in which case \re{YMsph} is indeed general enough. Another difference of
$d>4$ versus $d=4$ is that in the latter case there is a duality between
the electric and magnetic fields while in our case this duality is absent.
Thus we do not seek to construct dyonic particle like solutions and have
set $A_0=0$ in \re{YMsph}.

For the system \re{einstYM} under consideration, namely that with
$P_1=P_2=2$, the variational equations for the YM and gravitational
fields are, respectively
\be
\tau_1D_{\mu}\left(e\ F^{\mu\nu}\right)+
\frac12\tau_2\{F_{\rho\si},D_{\mu}\left(e\ F^{\mu\nu\rho\si}\right)\}=0
\label{varYM}
\ee
\be
\ka_1G_{(1)}{}_{\mu}^a+\ka_2G_{(2)}{}_{\mu}^a\ =\
\tau_1T_{(1)}{}_{\mu}^a+\frac{1}{3!}\tau_2T_{(2)}{}_{\mu}^a\ .
\label{vargrav}
\ee
It is easy to systematically extend \re{varYM} and \re{vargrav} to the
highest values of $P_1$ and $P_2$ respectively, in any dimension $d$.

By virtue of the spherically symmetric YM Ansatz \re{YMsph} and the
metric Ansatz \re{metric}, it follows that the Gauss Law equation, namely
the $\nu=0$ (or $\nu=t$) component of \re{varYM} is identically
satisfied. Amongst the $\nu=i$ components, the $\nu=r$ also trivialise and
the $d-2$ equations pertaining to the $\nu=\theta_{\al}$ components
all coincide to give only one equation for \re{YMsph}. We list explicitly
the two terms in \re{YMsph}
\bea
D_{\mu}\left(e\ F^{\mu\theta_{\al}}\right)&=&
\left(r^{d-4}\sqrt{AB}w'\right)'
-(d-3)r^{d-6}\sqrt{\frac{A}{B}}(w^2-1)w \label{DYM1} \\
\{F_{\rho\si},D_{\mu}\left(e\ F^{\mu\theta_{\al}\rho\si}\right)\}&=&3
(d-3)(d-4)(w^2-1)\Bigg(\left(r^{d-8}\sqrt{AB}(w^2-1)w'\right)'\nonumber \\
&-&(d-5)r^{d-10}\sqrt{\frac{A}{B}}(w^2-1)^2w\Bigg) \label{DYM2}
\eea
for all $\al=1,2,...(d-2)$.

It follows from the metric Ansatz \re{metric} that the Einstein tensors
defined in \re{peinstein} have nonvanishing components $G_{(p)}{}_t^t$,
$G_{(p)}{}_r^r$ and $G_{(p)}{}_{\theta_{\al}}^{\theta_{\al}}$
($\al=1,2,...,(d-2)$), which can readily be calculated for any $p$. Here
we list explicitly the two cases of interest, $p=1$ and $p=2$,
repectively,
\bea
G_{(1)}{}_t^t&=&\frac{1}{2r}(d-2)\left[B'-\frac{1}{r}(d-3)
(1-B)\right]\label{ein1tt}\\
G_{(1)}{}_r^r&=&\frac{1}{2r}(d-2)\left[\frac{BA'}{A}
-\frac{1}{r}(d-3)(1-B)\right]\label{ein1rr}\\
G_{(1)}{}_{\theta_{\al}}^{\theta_{\al}}&=&\frac12\Bigg[\left(
\frac{BA''}{A}+\frac{1}{2}\left(\frac{B}{A}\right)'A'\right)
-\frac{1}{r^2}(d-3)(d-4)(1-B)\nonumber \\
&+&\frac{1}{r}(d-3)\frac{(AB)'}{A}\Bigg]\ ,\label{ein1aa}
\eea
and
\bea
G_{(2)}{}_t^t&=&\frac{1}{4r^3}(d-2)(d-3)(d-4)(1-B)
\Bigg[B'-\frac{1}{2r}(d-5)(1-B)\Bigg]\label{ein2tt} \\
G_{(2)}{}_r^r&=&-\frac{1}{4r^3}(d-2)(d-3)(d-4)(1-B)
\Bigg[\frac{BA'}{A}-\frac{1}{2r}(d-5)(1-B)\Bigg]\label{ein2rr} \\
G_{(2)}{}_{\theta_{\al}}^{\theta_{\al}}&=&\frac{1}{4r^2}(d-3)(d-4)
\Bigg[(1-B)\left(\frac{BA''}{A}+\frac{1}{2}
\left(\frac{B}{A}\right)'A'\right)-\left(\frac{BA'}{A}\right)B'\nonumber\\
&+&\frac{1}{r}(d-5)(1-B)\frac{(AB)'}{A}
-\frac{1}{2r^2}(d-5)(d-6)(1-B)^2\Bigg]\label{ein2aa}\ .
\eea
It follows from the Ans\"atze \re{metric} and \re{YMsph} that the stress
tensors defined in \re{pstress} have nonvanishing components
$T_{(p)}{}_t^t$, $T_{(p)}{}_r^r$ and
$T_{(p)}{}_{\theta_{\al}}^{\theta_{\al}}$ ($\al=1,2,...,(d-2)$), which
can readily be calculated for any $p$. Here we list explicitly the two
cases of interest, $p=1$ and $p=2$, repectively,
\bea
T_{(1)}{}_t^t&=&-\frac{1}{2^4r^2}\ n_d\ (d-2)
\left[\ \ 2Bw'^2+(d-3)\left(\frac{w^2-1}{r}\right)^2\right]\label{T1tt}\\
T_{(1)}{}_r^r&=&-\frac{1}{2^4r^2}\ n_d\ (d-2)
\left[-2Bw'^2+(d-3)\left(\frac{w^2-1}{r}\right)^2\right]\label{T1rr}\\
T_{(1)}{}_{\theta_{\al}}^{\theta_{\al}}&=&-\frac{1}{2^4r^2}\ n_d\
\left[2(d-4)Bw'^2
+(d-3)(d-6)\left(\frac{w^2-1}{r}\right)^2\right]\label{T1aa}\ ,
\eea
and
\bea
T_{(2)}{}_t^t&=&-\frac{(3!)^2}{2^7r^4}\ n_d\frac{(d-2)!}{(d-5)!}
\left(\frac{w^2-1}{r}\right)^2
\left[\ \ 4Bw'^2+(d-5)\left(\frac{w^2-1}{r}\right)^2\right]\label{T2tt}\\
T_{(2)}{}_r^r&=&-\frac{(3!)^2}{2^7r^4}\ n_d\frac{(d-2)!}{(d-5)!}
\left(\frac{w^2-1}{r}\right)^2
\left[-4Bw'^2+(d-5)\left(\frac{w^2-1}{r}\right)^2\right]\label{T2rr}\\
T_{(2)}{}_{\theta_{\al}}^{\theta_{\al}}&=&-\frac{(3!)^2}{2^7r^4}\
n_d\frac{(d-3)!}{(d-5)!}\left(\frac{w^2-1}{r}\right)^2\left[4(d-8)Bw'^2
+(d-5)(d-10)\left(\frac{w^2-1}{r}\right)^2\right]\label{T2aa}\ . 
\eea
\re{T1aa} and \re{T2aa} are valid for all $\al=1,2,...,(d-2)$. The
constants ${n_d=\mbox Tr}\ \eins$, where the dimensionality of the unit
matrix is determined by the chiral representations appearing in
\re{YMsph}.

Substituting \re{DYM1}-\re{DYM2} in \re{varYM} results in one equation,
and substituting \re{ein1tt}-\re{T2aa} in \re{vargrav} results in three
equations. Thus we have 4 ordinary differential equations for the 3
radial functions $A(r),\ B(r)$ and $w(r)$. This necessitates
the verification of the consistency of these 4 equations to ensure that
the system is not overdetermined. We have carried out this consistency
check in the case at hand with $P_1=P_2=2$ in \re{gravhier} in
\re{YMhier} respectively, for any $\ka_1$, $\ka_2$, $\tau_1$, $\tau_2$
and in any dimension $d$. We found that the components
$({\theta_{\al}},{\theta_{\al}})$ of \re{vargrav},
\be
\ka_1G_{(1)}{}_{\theta_{\al}}^{\theta_{\al}}+
\ka_2G_{(2)}{}_{\theta_{\al}}^{\theta_{\al}}\ =\
\tau_1T_{(1)}{}_{\theta_{\al}}^{\theta_{\al}}+
\frac{1}{3!}\tau_2T_{(2)}{}_{\theta_{\al}}^{\theta_{\al}}\ .
\label{vargravth}
\ee
are identically satisfied by the two other Einstein equations
\bea
\ka_1G_{(1)}{}_t^t+\ka_2G_{(2)}{}_t^t&=&
\tau_1T_{(1)}{}_t^t+\frac{1}{3!}\tau_2T_{(2)}{}_t^t\label{vargravt}\\
\ka_1G_{(1)}{}_r^r+\ka_2G_{(2)}{}_r^r&=&
\tau_1T_{(1)}{}_r^r+\frac{1}{3!}\tau_2T_{(2)}{}_r^r\label{vargravr}
\eea
together with the YM equation \re{varYM}.
Thus the 3 effective constraints on the 3
radial functions are the 2 equations \re{vargravt}-\re{vargravr} plus the
YM equation \re{varYM}. This check of consistency extends by induction to
the case of arbitrary $P_1\ ,\ P_2\ $.

Finally we express these 3 equations employing the same notation as
\cite{BK} and \cite{VG}, namely
\be
\label{notation}
A(r)\ =\ \si^2(r)\ N(r)\quad ,\quad B(r)\ =\ N(r)\ .
\ee
For the functions $w(r)$, $N(r)$ and $\si(r)$, we have
\bea
&2\tau_1&\left(\left(r^{d-4}\si Nw'\right)'
-(d-3)r^{d-6}\si(w^2-1)w\right)+\nonumber \\
+&3\tau_2&(d-3)(d-4)(w^2-1)\left(
\left(r^{d-8}\si N(w^2-1)w'\right)'
-(d-5)r^{d-10}\si(w^2-1)^2w\right)=0
\label{YM12eq}
\eea
\bea
m'&=&\frac18r^{d-4}\Bigg(\tau_1\left[Nw'^2
+\frac12(d-3)\left(\frac{w^2-1}{r}\right)^2\right] \nonumber \\
&+&\frac32\frac{\tau_2}{r^2}(d-3)(d-4)\left(\frac{w^2-1}{r}\right)^2
\left[Nw'^2+\frac14(d-5)\left(\frac{w^2-1}{r}\right)^2\right]\Bigg)
\label{meq}
\eea
\be
\left[\ka_1+\frac{\ka_2}{2r^2}(d-3)(d-4)(1-N)\right]\left(
\frac{\si'}{\si}\right)=\frac{n_d}{8r}
\left[\tau_1+\frac32\frac{\tau_2}{r^2}(d-3)(d-4)
\left(\frac{w^2-1}{r}\right)^2\right]w'^2\ .
\label{sigeq}
\ee
In \re{meq}, the function $m(r)$ is defined as
\be
\label{mdef}
m(r)=n_d^{-1}[\ka_1r^{d-3}(1-N)+\frac14\ka_2r^{d-5}(1-N)^2]\ ,
\ee
which for $\ka_2=0$ and with $d=4$ reduces to the definition of the
corresponding function $m(r)$ used in \cite{BK,VG}.
For $\ka_2=\tau_2==0$ and with $d=4$, equations \re{YM12eq}-\re{sigeq}
coincide with the ordinary diffrential equations of \cite{BK,VG}.

\section{Numerical results}
\label{numerics}
In this section we present the results we obtained for the two
cases corresponding to $p=1$ and $p=2$ gravity and for the corresponding
space-time dimensions $d=6$ and $d=8$ where solutions of the extendend
Yang-Mills equations (consisting of the superposition of the $p=1$ and
$p=2$ members of the YM hierarchy) exist even in the absence of gravity.
(The cases of odd values of $d$ will be addressed elsewhere).
When the two mixing parameters $\tau_1, \tau_2$ do not vanish
we can set them equal to particular values without losing
generality. This can be done by an appropriate
rescaling of the overal lagrangian density and of the radial variable.
So we take advantage of this freedom and choose $\tau_1 = 1 ,
\tau_2 = 1/3$ (this simplifies the numerical coefficients in the
equations). The limit $\tau_1\to 0$ is addressed separately.

\subsection{Boundary conditions}
We have solved the above equations with the appropriate boundary
conditions for the radial functions $m(r)$, $\si(r)$ and $w(r)$ which
guarantee the solution to be regular at the origin and to have finite
energy. These boundary values are
\be
\label{origin}
m(0)=0\ \ ,\qquad w(0)=1\ \ ,
\ee
at the origin, and
\be
\label{infty}
\lim_{r\to\infty}\si(r)=1\ \ ,\qquad \lim_{r\to\infty}w(r)=-1\ .
\ee
at infinity. The condition on $\si$ results in Minkowskian metric
asymptotically.

Further analysis of the YM equation \re{YM12eq} yields
\be
\label{YMasym}
w(r\rightarrow \infty) = -1 + \frac{C}{r^{d-3}}
\ee

\subsection{$p=1$ gravity: Pure Einstein-Hilbert}
We define here  $\alpha^2  \equiv n_d / (8 \kappa_1) $
so that  $\alpha^2 = 0$ corresponds to the decoupling of gravity
(and therefore to a flat space).

\subsubsection{$d=6$ case}
The flat solution corresponding to the flat case \cite{butc}
is characterized by  an energy $M \approx 7.096$ (in the conventions
used). This solution is smoothly deformed by gravity and exists up to
a critical value $\alpha^2_m \approx 0.1266$.

When $\alpha^2$ increases, the  mass of the gravitating solution decreases
from $M \approx 7.096$ (flat case) down to $M \approx 6.56$;
the function $N(x)$ develops a local
minimum, say $N_m$ which becomes deeper while gravity becomes stronger
and the value $\sigma(0)$ decreases from one.
These quantities are plotted in Fig.1 and the mass is reported
on Fig. 2

Our numerical analysis strongly indicates that a second branch of
regular solutions exists, also termin`ating at $\alpha^2 = \alpha_m^2$.

On this second branch the values $\sigma(0)$
and $N_m$ increase monotonically with $\alpha^2$.
The quantity $\sigma(0)$ reaches the value zero
 $\alpha^2 \rightarrow \alpha^2_c$ , with $\alpha^2 \approx 0.07$,
as illustrated by Fig. 1.

Accordingly, the metric becomes
singular in this limit and no solutions
seem to exist on this branch for lower values of $\alpha^2$.

The second branch of solutions therefore exist for
$\alpha_c < \alpha^2 \leq < \alpha_c^2$. The mass of the solution
of the second branch is larger than the corresponding one on the
first branch, as illustrated by Fig. 2.

The minimal value of $N(r)$ stays well above zero ($N_{min}\approx 0.17$)
so that no horizon is approached. The situation contrasts with the case
of the gravitating monopole in 3+1 dimensions \cite{w,bfm}. In
both cases (the present and that of \cite{w,bfm}) there occur second
branches of the solutions after critical values the respective parameters
($\al^2 in our case$), these bifurcations are quite different in the two
cases. 
The gravitating monopole bifurcates into a Reissner-Nordsrtom
black hole solution while the solutions studied here terminate
into a solution presenting a singularity at the origin. Several
components of the Riemann tensor are indeed polynomials is the field
$\sigma(r)$.

For scaling reasons, the parameter $\tau_2$ cannot vanish for
finite energy solutions to exist.
The numerical study of the equation with a varying $\tau_1$
(with $\alpha^2$ fixed) reveals
that, in the limit $\tau_1 \rightarrow 0$, the solutions of the
first branch  converge to the vacuum solution
(with mass equal to zero and with $w(x)=1$)
When the solutions on the second branch are examined in the same limit,
it appears that they stop to exist
at a critical value of $\tau_1$ where the value $\sigma(0)$
becomes zero. In other words, it also terminates into a singular solution
for a finite value of $\tau_1$ and no regular solution seems to exist
for $\tau_1 = 0$.

\subsubsection{$d=8$ case}
The scenario is qualitatively similar to the one of the $d=6$ case.
Here the flat solution has $M \approx 25.6$, it gets deformed by
gravity up to $\alpha^2 \approx 0.0022$ and the function $\sigma(r)$
is such that $\sigma(0)$ decreases with $\alpha^2$. At the same time
$N(r)$ develops a local minimum which becomes deeper. Another branch
of solution exists where
both $\sigma(0)$ and $N_{min}$ are lower than their conterparts on
the first branch. These relevant quantities are plotted in Figs.
3 and 4. On Fig. 5 we present the profiles of the metric functions
for several values of the parameters.

The numerical analysis suggests that this second branch persists
up to $\alpha^2 = 0$ and that in the limit,
$\lim_{\alpha^2 \rightarrow 0}\sigma(0) = 0$.
As far as our numerical analysis indicates, the value $N_m$
tends to a finite value in this limit so that there occurs no horizon.

We further investigated the two branches of solutions in the limit
$\tau_1 \rightarrow 0$. It appears that the first branch solution
tends to the vacuum while the second one tends to a non trivial
solution with a finite mass.

\subsection{$p=2$ gravity: Pure Gauss-Bonnet}
This case is not of any particular physical relevance but it is
interesting to learn the peculiarities of the pure $p=2$ gravity
especially when the YM term is also the pure $p=2$ member of the YM
hierarchy in $d=8$. Such a
system supports a self-dual YM solution on double--self-dual gravitational
background~\cite{BCT} when the metric has Euclidean signature, so it is
interesting to contrast with the case of Minkowskian signature at hand.

\subsubsection{$d=6$ case}
We define the constant $\beta^4  \equiv 4 n_d / \kappa_2 $ in this case.
Again the flat solution gets deformed by the gravitating term
and a gravitating solution is formed, characterized by
$\sigma(0), N_m$ (see Fig. 6) and by the mass (see Fig. 7).

While both $\sigma(0), N_m$ decrease monotonically as functions of
$\beta^2$, the mass first decreases, reaches a minimum for
$\beta^2 \approx 0.75$ and
then starts to increase very rapidly with $\beta^2$.

In the present case, it is very likely
that the gravitating solution   exists up to a maximal value
of $\beta^2$, say $\beta^2_m$ and that, when this limit is approached,
$N_m$ tends to zero, forming an horizon, while $M$ tends to infinity.
(Incidentally, $r_0$ the value of the radial variable where $w(r)$
crosses zero, exhibits the same behaviour).
The numerical integration of the equations for $\beta^2 > 0.87$ becomes
difficult and we could not evaluate $\beta^2$ with a good accuracy.

Denoting  $r_m$ the value of $r$ where the function $N(r)$ attains its
minimum, we observe that
the solutions possess a completely different behaviour in the regions
$r > r_m$ and $r < r_m$.

The functions $m(r), \sigma(r), w(r)$ vary essentially on the interval
  $[0, r_m]$ and attain their asymptotic values $m(\infty), 1 , -1$
for $r \simeq r_m$.
They are essentially constant on the interval $ r \in ]r_m, \infty[$.
This feature is illustrated in  Fig. 8.

Our belief that the solution stops at some maximal value of $\beta^2$
is reinforced by the fact that our attempts to produce a second branch
of solutions turned out to be unsuccessful.

A peculiarity of the equations is that
fixing $\beta^2$ and $\tau_1$ and varying $\tau_2$, is equivalent
to a rescaling of the radial variable $r$.
Fixing $\beta^2$ and decreasing $\tau_1\to 0$ the numerical results
strongly suggest that the solution tends to the vacuum.

\subsubsection{$d=8$ case}
The pattern of solutions is similar to the ones in the $d=6$ case.
Here the mass attains its minimum at $\beta^2 \approx 0.55$ and
again the gravitating solutions seem to exist up to $\beta^2 \approx 1.2$.
The relevant quantities characterizing the solutions are plotted
in Figs. 9 and 10.
The behaviour of the solutions for $\beta^2$ approaching the critical
value is illustrated on Fig. 11.

The absence of a solution for the model consisting purely of the $p=2$
members of both gravitational and YM hierarchies for Minkowskian
signature $d=8$ at hand, contrasts with the selfdual (black hole)
solutions~\cite{BCT} in the case of Euclidean signature. (Note that the
$p=1$ counterpart in Minkowskian $d=4$ exists, namely the solutions
of \cite{BK}, along with the selfdual solutions.) It would be interesting
to find out if the situation is different for Minkowskian signature
black hole solutions in $d=8$.

\subsection{$p=1 + p=2$ gravity: Superposed Einstein and Gauss-Bonnet}
We finally considered the equations for the case where the
$p=1$ and $p=2$ gravities are superposed. We limited our analysis
to the case where the $p=2$ gravity coupling is slightly smaller
(although not too small in order to see the  effect of the
superposition) than the $p=1$ parameter.
To fix the idea, we choose $ \kappa_2 = (1/2) \kappa_1$.
Our conclusion is that for $d=6$ and $d=8$ the qualitative properties
of the solutions corresponding to p=1 are conserved, namely
two branches of solutions are found.
The maximal value $\alpha^2$ increases slightly in the mixed case;
for instance $\alpha_m^2 \approx 0.004$ for our choice of the parameter
in the $d=8$ case.

\section{Conclusions}
\label{conclusions}
It appears that the difference of the solutions that we have
constructed, from the those given in~\cite{BK} for the
usual EYM system, arises due to the presence of the
additional dimensional constants $\ka_2$ and $\tau_2$. These constants
result in the presence of a nontrivial parameter in the equations,
for instance the one we denoted as $\al^2$, which contains the
gravitational coupling strength. By varying this parameter we were able
to construct one or several branches of solutions for the cases
considered. In this respect the system of equations studied is analogous
to that of the gravitating monopole~\cite{bfm} where the VEV of the Higgs
fields also gives rise to an additional dimensional parameter.

In this prelimary study, we have restricted ourselves to the construction
of the simplest static spherically symmetric (in $d-1$ spacelike
dimensions) solutions. Natural extensions of the present work include
{\it (i)} the construction of solutions with {\em several nodes} in the
YM function $w(r)$,  and {\it (ii)} the construction of the black hole
counterparts of our particle-like solutions. In the latter case these
pertain to the fully interacting version of the {\em fixed gravitational
background} solutions discussed in~\cite{BCT}. Another extension,
{\it (iii)}, concerns the deformation of our solutions due to the
inclusion of a cosmological constant~\cite{H}. Finally, {\it (iv)} the
stability analysis of our solutions can be carried out.

\medskip
\medskip

\noindent
{\bf Acknowledgements} This work was supported by an Enterprise--Ireland
grant for International Collaboration project IC/2001/073.


\begin{small}

\centerline{Figure Captions}
\begin{itemize}

\item [Figure 1]
The minimal value of the function $N$ and the value 
$\si(0)$ are plotted as functions of $\alpha^2$ for $p=1$ gravity
in $d=6$.
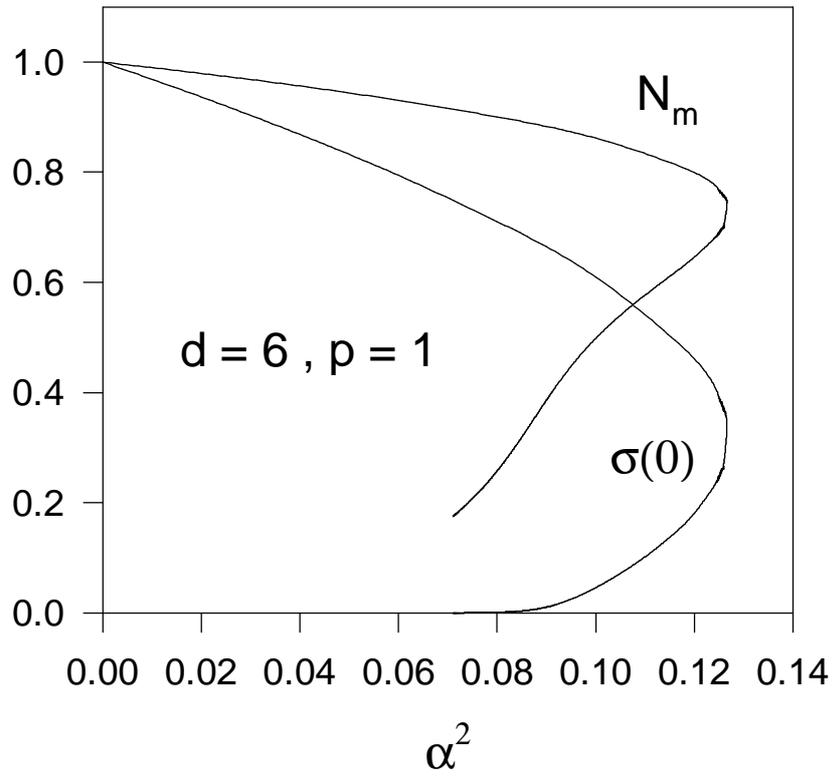
\item [Figure 2]
The mass of the solutions for $p=1$ gravity in $d=6$.
\item [Figure 3]
{\it Idem} Fig. 1 in $d=8$.
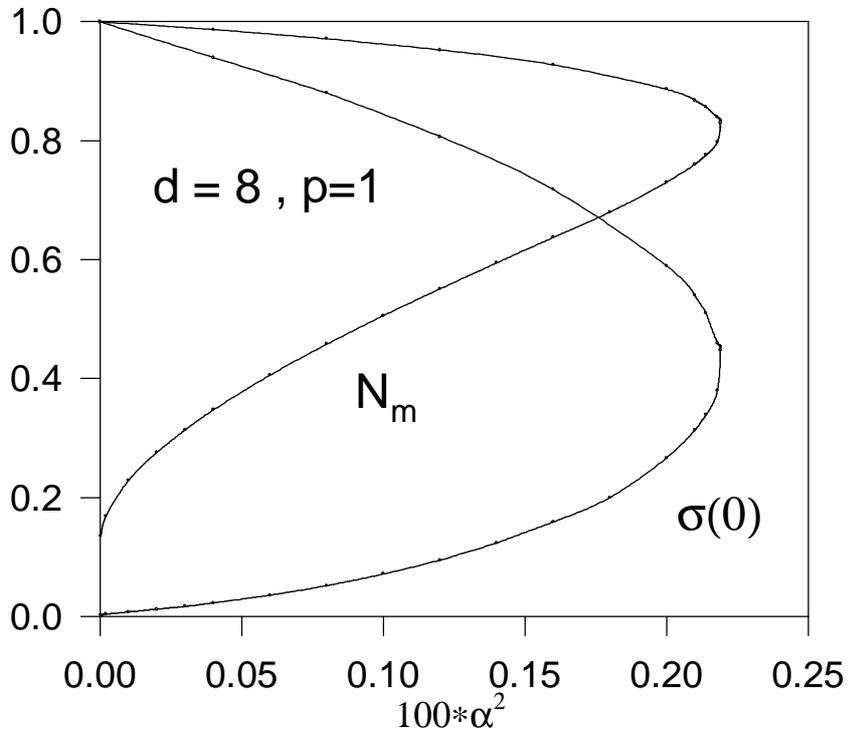
\item [Figure 4]
{\it Idem} Fig. 2 in $d=8$.
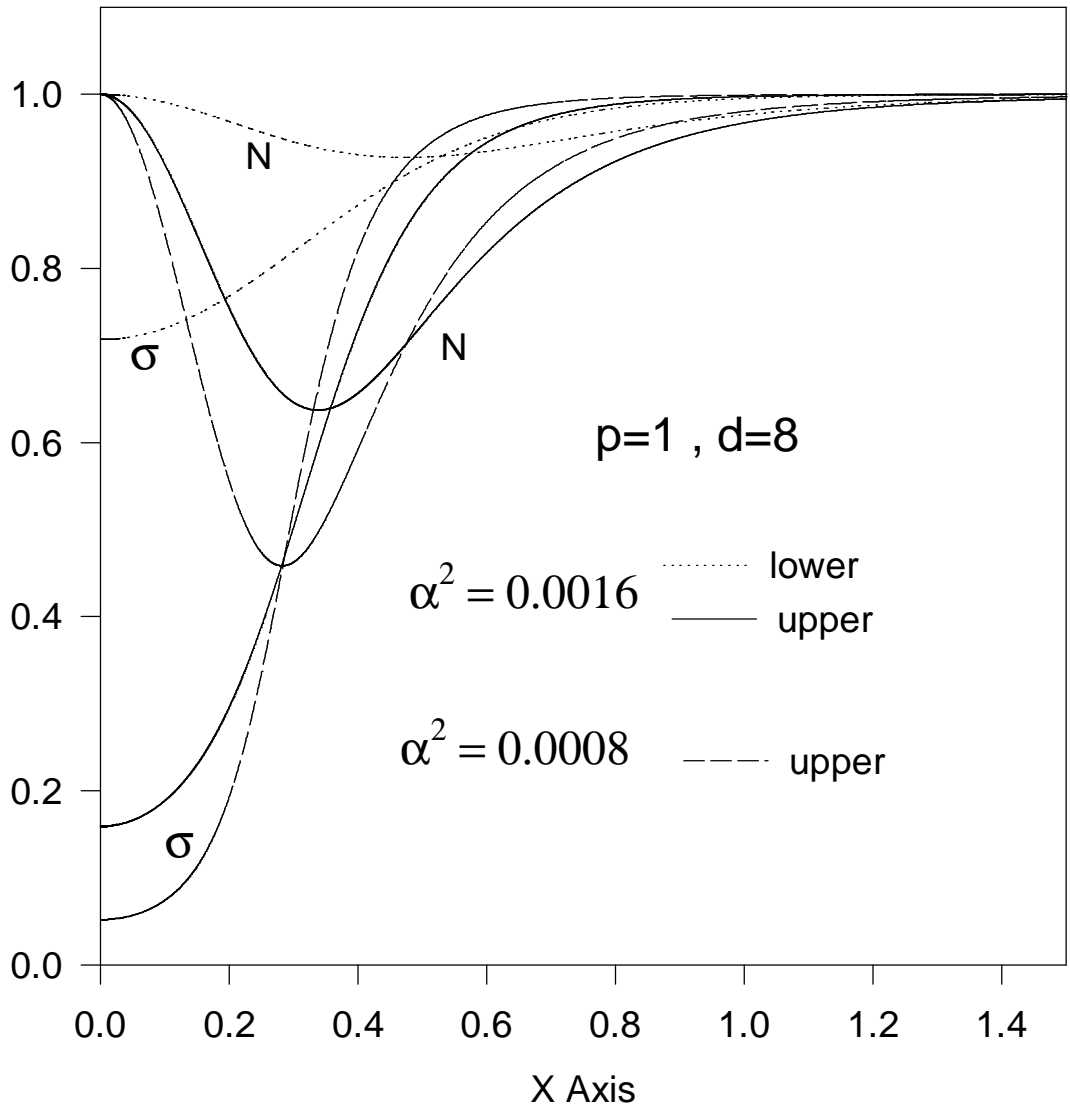
\item [Figure 5]
The profiles for the metric functions $N(r)$, $\si(r)$
are presented for different values of $\alpha^2$ for the
two branches for $p=1$ gravity in $d=8$. The lower branch plot of the
$\alpha^2 = 0.0008$ case is not reported because it is very
close to the flat solution.
{\it Idem} [Figure 6]
Idem Fig. 1 for $p=2$ gravity in $d=6$.
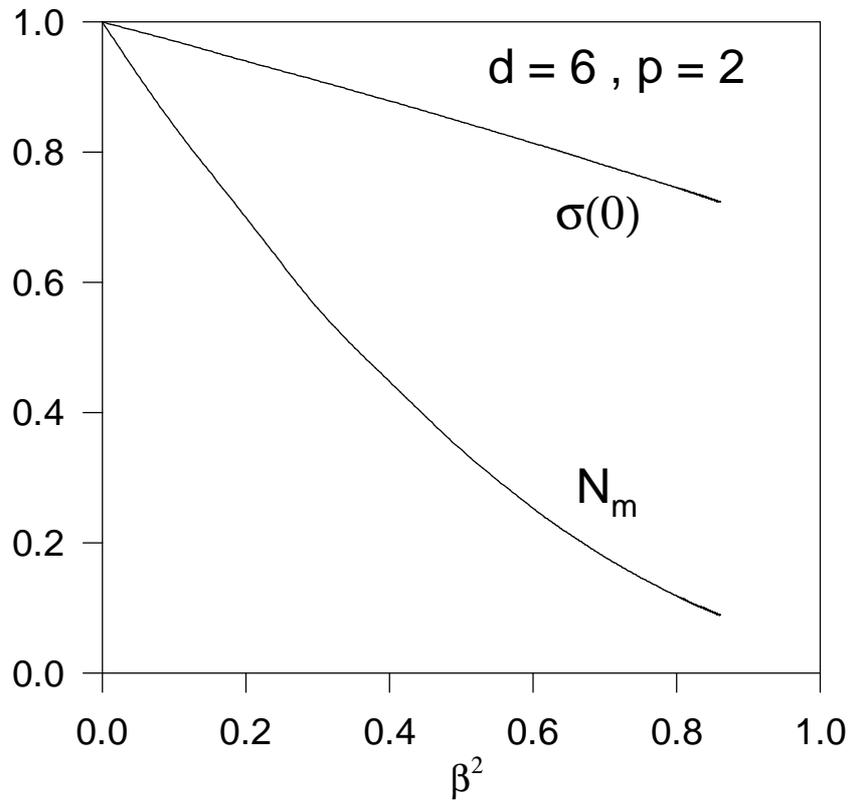
\item [Figure 7]
{\it Idem} Fig. 2 for $p=2$ gravity in $d=6$, 
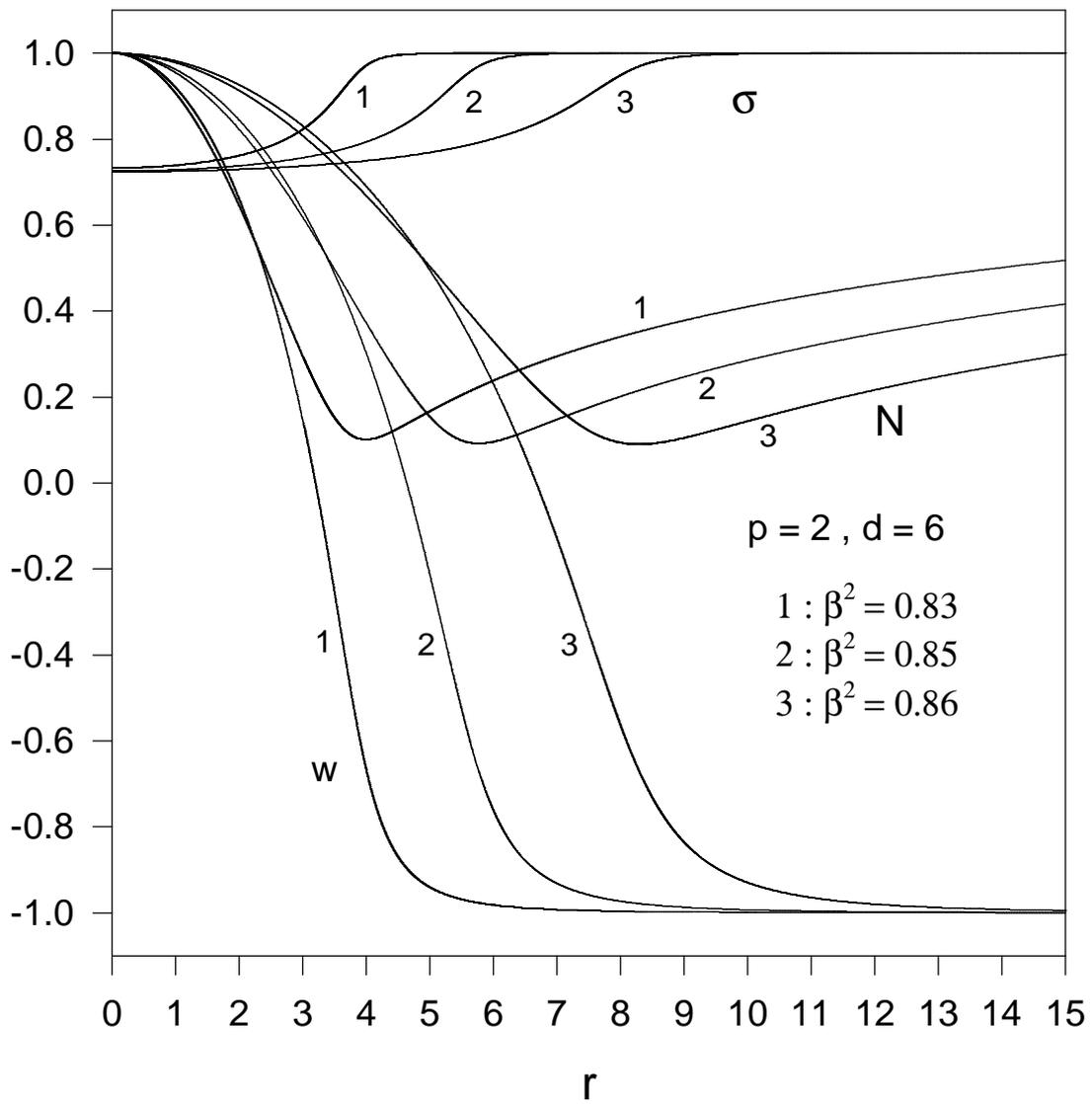
\item [Figure 8]
The evolutions of $N(r), \si(r), w(r)$,
at the approach of the critical value for $p=2$ gravity in $d=6$.
\item [Figure 9]
{\it Idem} Fig. 1 for $p=2$ gravity in $d=8$.
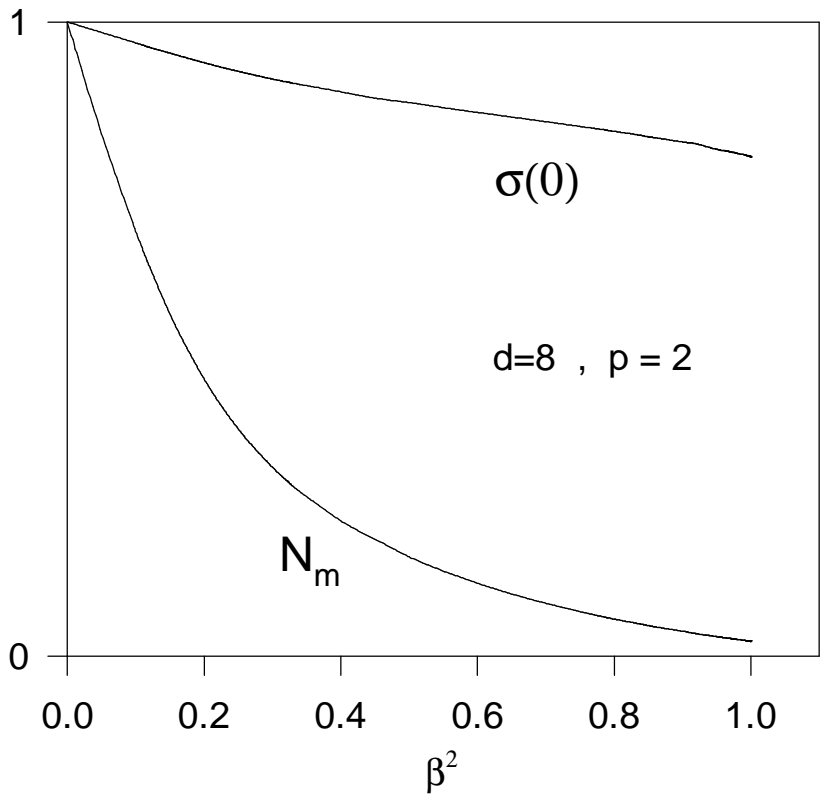
\item [Figure 10]
{\it Idem} Fig. 2 for $p=2$ gravity in $d=8$.
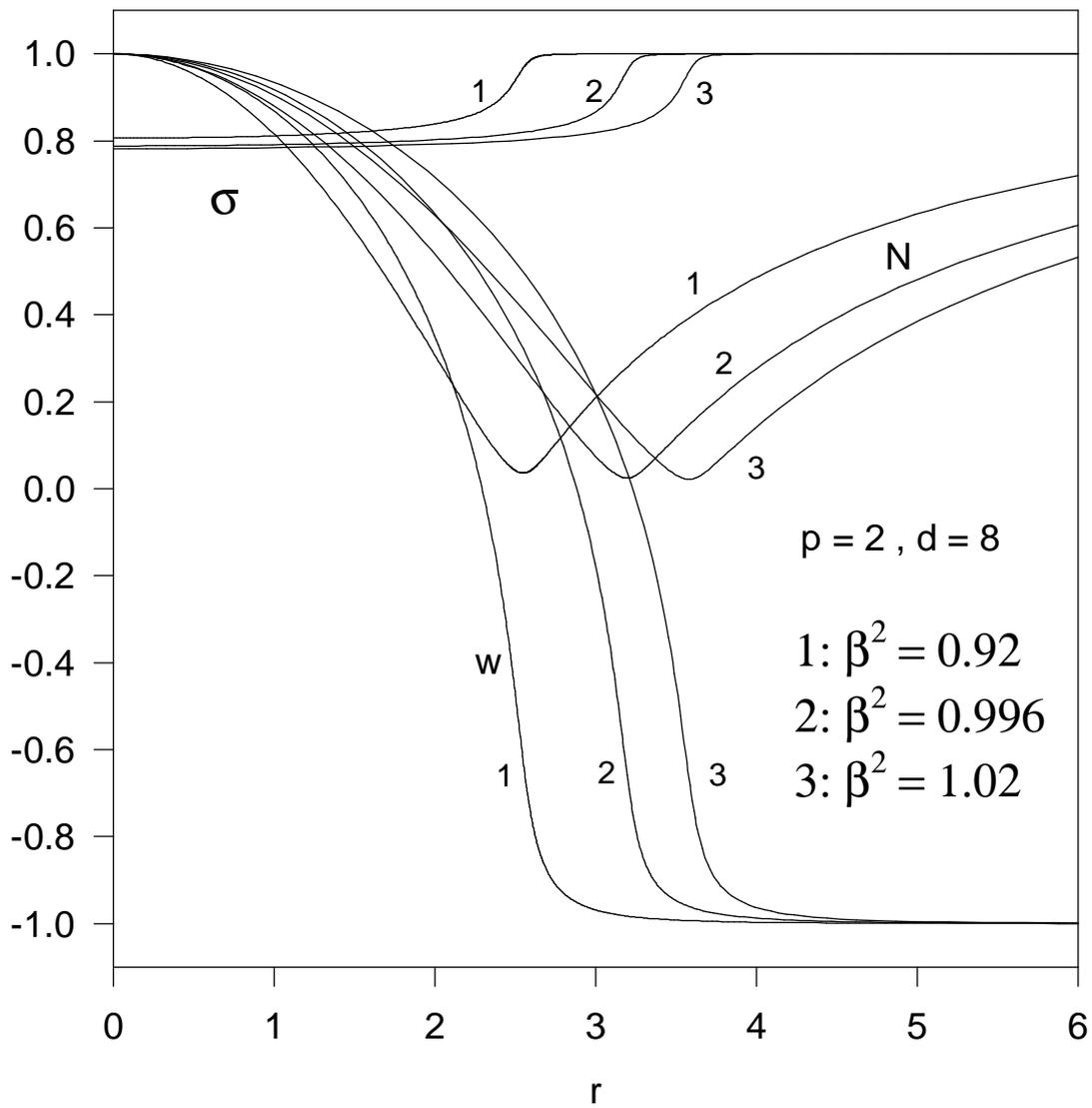
\item [Figure 11]
{\it Idem} Fig. 8 for $p=2$ gravity in $d=8$.
\end{itemize}

\end{small}

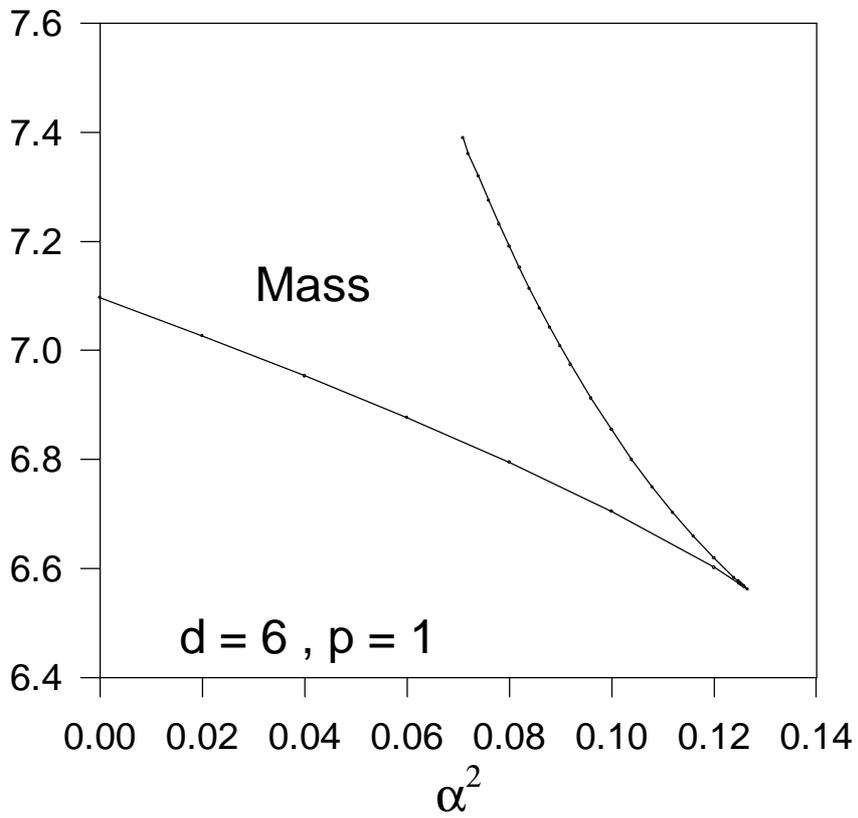
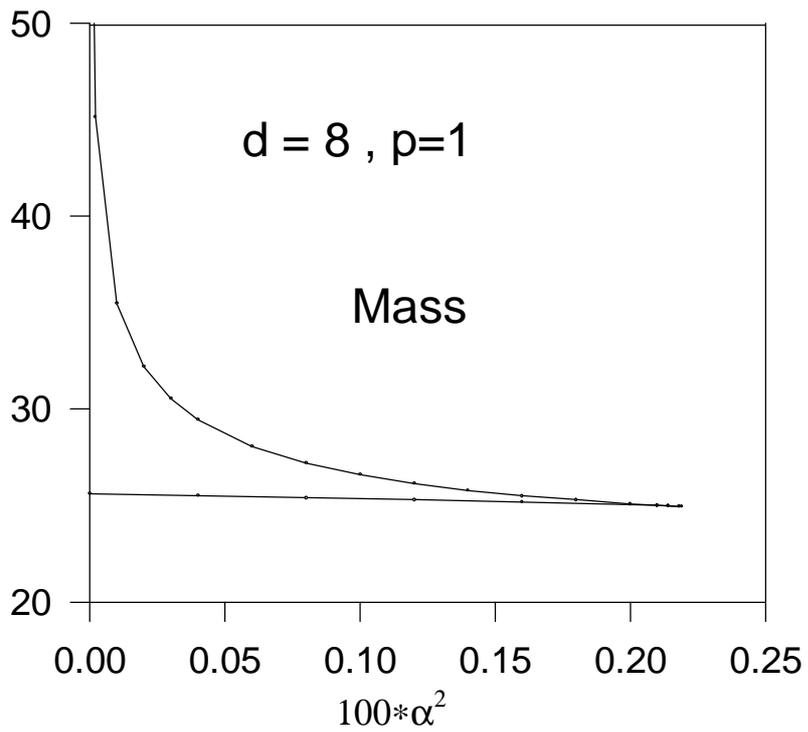
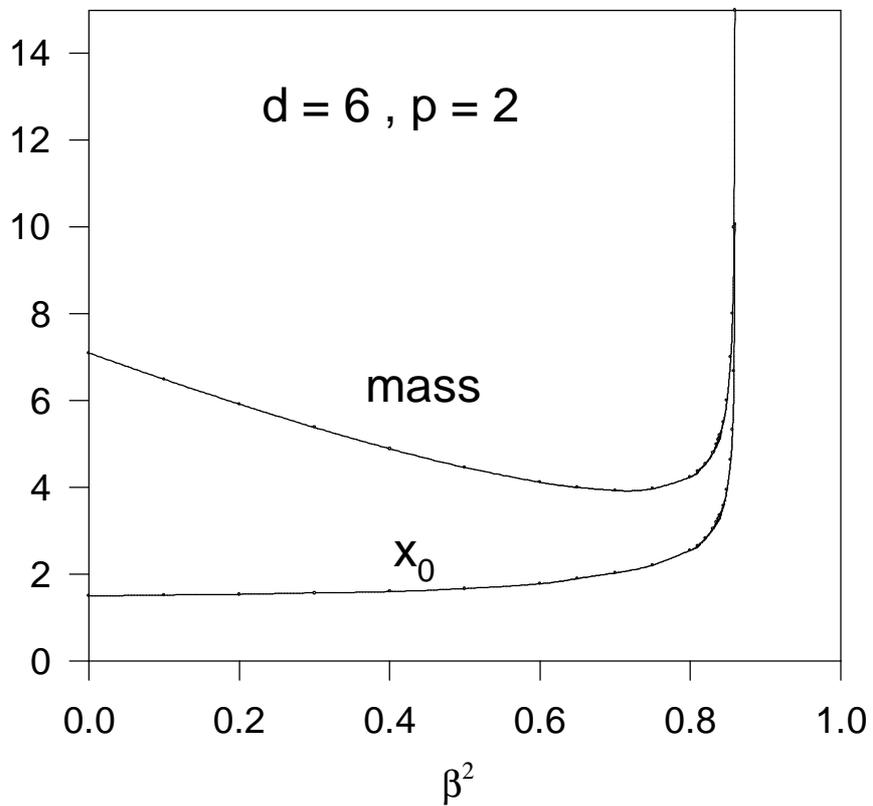
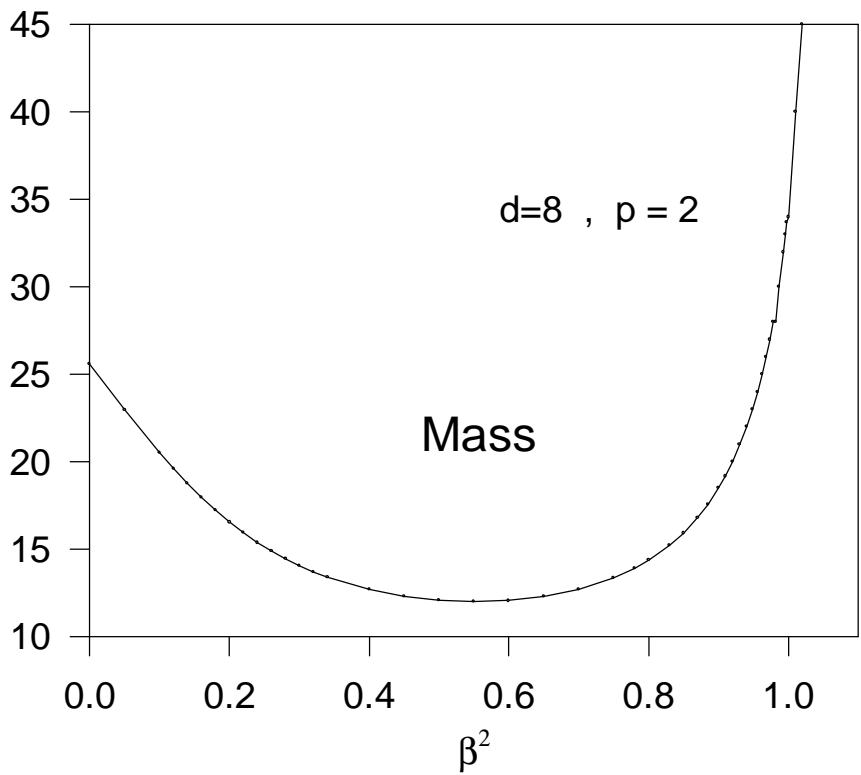
\end{document}